\documentstyle[psfig]{mn}
\newif\ifAMStwofonts
\newcommand{\lapp}{\mbox{\raisebox{-0.3em}{$\stackrel{\textstyle <}{\sim}$}}}
\newcommand{\gapp}{\mbox{\raisebox{-0.3em}{$\stackrel{\textstyle >}{\sim}$}}}

\title[H{\sc i} gas in rejuvenated radio galaxies]{H{\sc i} gas in rejuvenated radio galaxies: GMRT observations of 
                                                  the DDRG J1247+6723}
\author[D.J. Saikia et al.]
  {D.J. Saikia$\thanks{E-mail: djs@ncra.tifr.res.in (DJS), neeraj@ncra.tifr.res.in (NG), sckonar@ncra.tifr.res.in (CK)}$, 
  Neeraj Gupta and C. Konar \\
National Centre for Radio Astrophysics, TIFR, Pune University Campus, Post Bag 3, Pune 411 007, India}
\date{Accepted.    Received }

\pagerange{\pageref{firstpage}--\pageref{lastpage}}
\pubyear{  }

\begin{document}

\maketitle

\label{firstpage}

\begin{abstract}
We report the detection of H{\sc i} absorption towards the inner double of the double-double 
radio galaxy (DDRG) J1247+6723 with the Giant Metrewave Radio Telescope (GMRT). The inner 
double is a Giga-hertz peaked spectrum (GPS) source with a linear size of 14 pc while the overall
size defined by the outer double is 1195 kpc, making it a giant radio source.  
The absorption profile is well resolved and consists of a number of components on either
side of the optical systemic velocity.  The neutral hydrogen column density is estimated to be 
$N$(H{\sc i})=6.73$\times$10$^{20}$($T_s$/100)($f_c$/1.0) cm$^{-2}$, 
where $T_s$ and $f_c$ are the spin temperature and covering factor of the background source respectively. 
We explore any correlation between the occurrence of H{\sc i} absorption and rejuvenation of
radio activity and suggest that there could be a strong relationship between them.
\end{abstract}


\section{Introduction}
In the currently widely accepted paradigm, nuclear activity in 
active galactic nuclei (AGN) is believed
to be intimately related to the `feeding' of a supermassive black hole whose
mass ranges from $\sim$10$^6$ to 10$^{10}$ M$_\odot$. Such an active phase 
may be recurrent with  an average total timescale of the active phases being
$\sim$10$^8$ to 10$^{9}$ yr (cf. Marconi et al. 2004, and references therein). 
For the radio-loud AGN, an interesting way of probing their history and hence
episodic jet activity
is via the structural and spectral information of the lobes of extended 
radio emission (e.g.  Burns, Schwendeman \& White 1983; 
Burns, Feigelson \& Schreier 1983; van Breugel \& Fomalont 1984; Leahy, Pooley 
\& Riley 1986; Baum et al. 1990; Clarke, Burns \& Norman 1992; Junkes et al. 1993; 
Roettiger et al. 1994; Gizani \& Leahy 2003; Konar et al. 2006).
A very striking example of episodic jet activity is when a new pair of radio lobes
is seen closer to the nucleus before the `old' and more distant radio
lobes have faded (e.g. Subrahmanyan, Saripalli \& Hunstead 1996; Lara et al. 1999).
Such sources have been christened as `double-double' radio galaxies
(DDRGs) by Schoenmakers et al. (2000).
They proposed a relatively general definition of a DDRG as a double-double radio
galaxy consisting of a pair of double radio sources with a common centre.
Saikia, Konar \& Kulkarni (2006) reported the discovery of a new DDRG J0041+3224 and
compiled a sample of approximately a dozen such objects from 
the literature including 3C236 (Schilizzi et al. 2001) and J1247+6723 (Marecki et al. 2003;
Bondi et al. 2004). The inner doubles in these two sources are compact with sizes of 1.7 kpc and
14 pc respectively, and have been classified as a compact steep spectrum (CSS) and 
a Gigahertz peaked spectrum (GPS) source respectively. For the sample compiled by Saikia et al.
(2006) the size of the inner doubles ranges up to $\sim$650 kpc and has a median value of
$\sim$165 kpc, while that of the outer doubles range from $\sim$400 kpc to 4250 kpc and has
a median value of $\sim$1100 kpc.  

If the nuclear or jet activity is rejuvenated by a fresh supply of gas one might
be able to find evidence of this gas via H{\sc i} absorption towards the radio components
in the central regions of the host galaxy. It is interesting to note that the well-known
DDRG 3C236, which is the largest giant radio galaxy with a projected linear size of 4250 
kpc, shows evidence of star formation and H{\sc i} absorption against a lobe of the inner 
double (Conway \& Schilizzi 2000; Schilizzi et al. 2001; O'Dea et al. 2001).
Considering this, we observed the other DDRG J1247+6723, selected from the compilation of
Saikia et al. (2006), which has a compact inner double (Marecki et al. 2003;
Bondi et al. 2004) to investigate the occurrence of H{\sc i} absorption in this class of
objects. We describe some of the basic properties of J1247+6723 in Section 2, the observations
and analyses in Section 3 and present the results and discussions in Section 4. In Section 4
we also explore any possible relationship between detection of H{\sc i} absorption and evidence 
of rejuvenation of jet activity in powerful radio galaxies. The conclusions are summarised in Section 5.
  
\begin{figure}
\vbox{
  \psfig{file=J1247NVSS.PS,width=3.35in,angle=0}
   \psfig{file=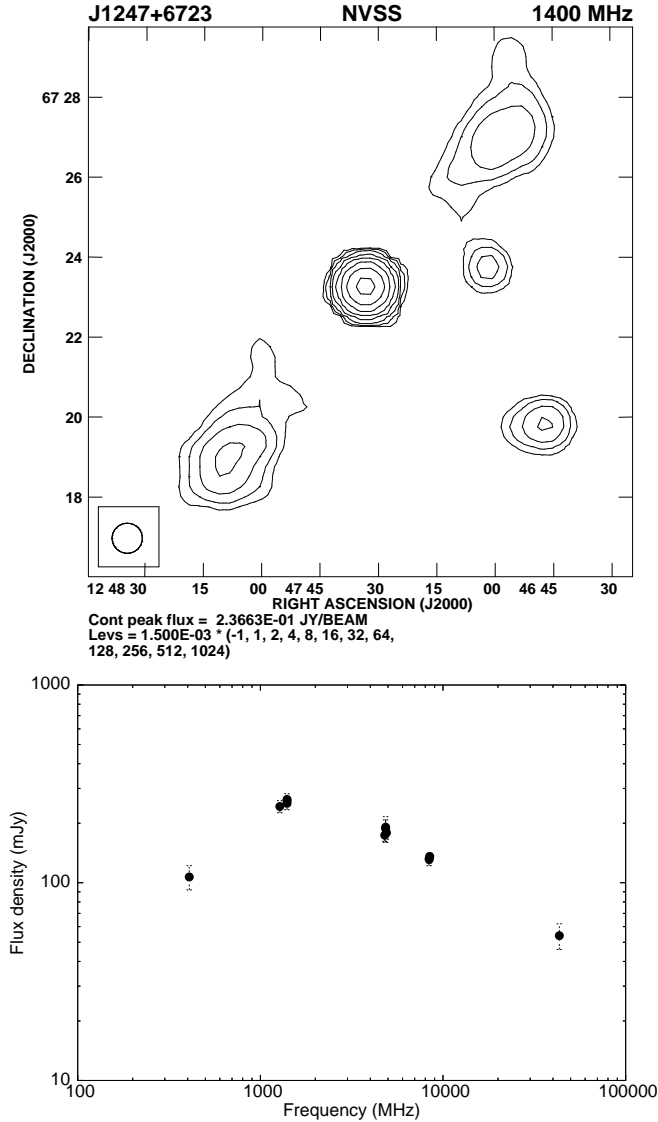,width=3.35in,angle=-90}
    }
\caption[]{The NVSS image of J1247+6723 with an angular resolution of
45 arcsec (upper panel) and the radio spectrum of the central component which gets
resolved into a compact inner double when observed with milliarcsec resolution (lower panel).
The restoring beam of the NVSS image is indicated by an ellipse. }
\end{figure}

\begin{table}
\caption{Flux densities of the inner double of 
J1247+6723 (VII Zw 485).}
\begin{center}
\begin{tabular}{|c|c|c|c|}
\hline
Frequency  &  Flux density   &  Error   &  References  \\
MHz    &      mJy        &   mJy    &         \\ 
\hline
408    &  107            &   15     &   1     \\  
1280   &  243            &   17     &   P     \\  
1400   &  262            &   08     &   2     \\  
1400   &  264            &   18     &   3     \\  
1400   &  252            &   18     &   4     \\  
1400   &  270            &   14     &   5     \\ 
4850   &  174            &   15     &   3,10  \\  
4850   &  188            &   20     &   6     \\ 
4850   &  191            &   17     &   7     \\  
4900   &  179            &   13     &   2     \\ 
8400   &  133            &    7     &   8     \\  
8400   &  131            &    9     &   3     \\  
8460   &  136            &    4     &   9     \\  
43340$^a$  &  54         &    8     &   9     \\ 
\hline
\end{tabular}

1: Marecki et al. 1999; 2: Lara et al. 2001; 3: March\~a et al. 2001; 
4: Caccianiga \& March\~a  2004; 5: our estimate from the NVSS image;  
6: Becker, White \& Edwards 1991; 7: Gregory \& Condon 1991;
8: Patnaik et al. 1992; 9: Dennett-Thorpe \& March\~a 2000; 
10: Gregory et al. 1996; P: present 
paper.  \\
$^a$ Extrapolated from the 8460 MHz flux density using the spectral
index given by Dennett-Thorpe \& March\~a (2000).
\end{center}
\label{}
\end{table}

\section{J1247+6723 (VII Zw 485)}
The radio source J1247+6723 is associated with the galaxy VII Zw 485 at a redshift of 
0.1073$\pm$0.0002 (Falco et al. 1998; de Vries et al. 2000; Goncalves et al. 2004). 
Spectroscopic observations of the galaxy show the presence of H$\alpha$, H$\beta$, 
[N{\sc ii}]$\lambda$6584, 6548  \AA~ and [S{\sc ii}]$\lambda$6717, 6731 \AA~ in emission leading
to its classification as a possible LINER (Goncalves et al. 2004 and references therein). 
A redshift of 0.1073 implies that the luminosity distance is 490.5 Mpc and 1 arcsec
corresponds to 1.939 kpc in a Universe with H$_\circ$=71 km s$^{-1}$ Mpc$^{-1}$, $\Omega_m$=0.27, 
$\Omega_\Lambda$=0.73 (Spergel et al. 2003).

The radio source consists of two lobes separated by $\sim$10.3 arcmin (1195 kpc) and a compact
radio core which is resolved into a small double with a separation of $\sim$7.2 milli arcsec
(14 pc). The large-scale radio structure has been imaged by Lara et al. (2001) while the compact
inner double, with both mini lobes having a steep high-frequency spectrum, has been observed by Marecki et al.
(2003) and Bondi et al. (2004). The NVSS (NRAO VLA Sky Survey) image of the source showing the
central component and the outer double lobes is presented in Fig. 1 (upper panel). The two sources 
south of the northern
lobe are likely to be unrelated sources. The total flux density of the source estimated
from this image is 379 mJy while that of the central component is 270 mJy using a two-dimensional
Gaussian fit. The flux densities of the central component which consists of the inner double are 
summarised in Table 1, while its spectrum showing that it is a 
GPS source is presented in Fig. 1 (lower panel). All the flux densities are consistent with the
scale of Baars et al. (1977). It is also clear from
the different measurements that the source does not exhibit evidence of any significant variability.
This along with its lack of radio polarization (Dennett-Thorpe \& March\~a 2000) is consistent with
its GPS status (cf. O'Dea 1998). The radio luminosity of the inner double at 1.4 GHz is
7.4$\times$10$^{24}$ W Hz$^{-1}$ while that of the outer double is 3.5$\times$10$^{24}$ W Hz$^{-1}$;
thus the inner double contributes $\sim$68 per cent of the total luminosity.
 
\section{Observations and analyses} 
We observed J1247+6723 with the GMRT to search for
associated 21-cm absorption towards the inner double. The source was observed on 2005 Dec 11
with a bandwidth of 8 MHz for $\sim$5.5 hours including calibration overheads.
In these observations  we made use of the new high-resolution mode of the
GMRT correlator software. This allows one to split the baseband bandwidth into 256 channels,
instead of the usual 128 channels.  This provided us a
spectral resolution of $\sim$7 km s$^{-1}$.
The local oscillator chain and FX correlator
system were tuned to centre the baseband bandwidth at 1282.76 MHz,
the redshifted 21-cm frequency corresponding to $z_{em}$=0.1073.  We observed the standard
flux density calibrator 3C286 every 3 hours to correct for the variations
in amplitudes and bandpass.  The compact radio source J1313+675 was also observed approximately
every 35 minutes for phase calibration of the array. The flux density of the phase calibrator 
was estimated to be 2.37$\pm$0.04 Jy. A total of $\sim$3.5 hours of data on source
were acquired in both the circular polarization channels RR and LL.

%%%%
The data were reduced in the standard way using Astronomical Image Processing System ({\tt AIPS})
package.  After the initial flagging or editing of bad data and calibration, source and
calibrator data were examined for baselines and timestamps affected by Radio Frequency
Interference (RFI).  These data were excluded from further analysis.  A continuum image
of the source was made using calibrated data averaged over 65 line-free channels.
Due to our interest in absorption towards the compact central source, the imaging
was done without any {\it uv} cut-off or tapering in the visibility plane. This provided us
with the highest possible resolution.
This image was then self-calibrated until a satisfactory map was obtained. The 
milliarcsec-scale inner double was seen as a single compact radio source. 
The self-calibration complex gains determined from this were applied to all the 256
frequency channels and continuum emission was subtracted from this visibility data cube using
the same map.  Spectra at peak position of the central source was extracted from
this cube. The whole process was done separately for Stokes RR and LL to check for consistency.
The two polarization channels were then combined to get the final Stokes I spectrum, which
was then shifted to the optical heliocentric frame.

\begin{figure}
\vbox{
    \psfig{file=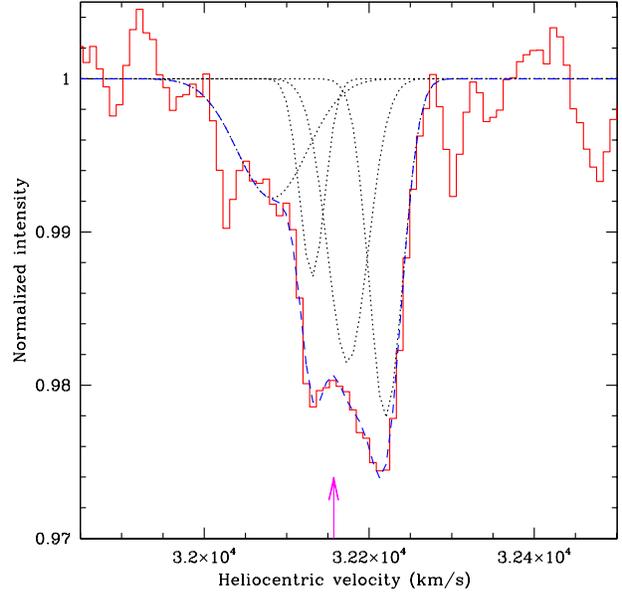,width=3.35in,angle=0}
    }
\caption[]{The H{\sc i} absorption spectrum (histogram) towards the compact inner double of the radio 
galaxy J1247+6723.  The spectrum has been smoothed using a 3-pixel wide boxcar filter.  The 
Gaussian components fitted to the absorption profile and the sum of these components i.e. 
the fit are plotted as dotted and dashed lines respectively.  The systemic velocity 
(32157 km s$^{-1}$) has been marked by an arrow.  
}
\end{figure}

\section{Results and discussion}

% %%%%%%% %% %%%%%%%%%
%
\begin{table}
\caption{Multiple Gaussian fit to the H{\sc i} absorption spectrum.}
\begin{center}
%\vspace{5mm}
\begin{tabular}{|c|l|l|c|c|}
\hline
Id. &   v$_{\rm{hel}}$ &  FWHM      & Frac. abs. & $N$(H{\sc i}) \\
no. &               &               & & 10$^{20}$($\frac{T_s}{100})(\frac{f_c}{1.0})^{-1}$  \\
    &   km s$^{-1}$ &   km s$^{-1}$ & & cm$^{-2}$ \\
\hline
1   &        32220  &    47.5(1.3)  &  0.022(0.001)  &    2.02(0.15)    \\                        
2   &        32174  &    57.8(6.1)  &  0.019(0.001)  &    2.12(0.34)    \\                              
3   &        32131  &    35.5(2.1)  &  0.013(0.001)  &    0.89(0.12)    \\                             
4   &        32083  &    99.6(8.0)  &  0.008(0.001)  &    1.54(0.32)    \\                              
\hline
\end{tabular}
\end{center}
\label{gauss}
\end{table}
 % %%%%%%%%%%%%%%%

Our GMRT image of the source with an angular resolution of
3.16$\times$2.21 arcsec$^2$ along a position angle of 157$^\circ$ and an rms
noise of 0.14 mJy beam$^{-1}$ detects only the central component which has a peak
brightness of 239 mJy beam$^{-1}$. A two dimensional Gaussian fit to the source
yields a peak brightness of 240 mJy beam$^{-1}$ and an integrated flux density of 
243 mJy beam$^{-1}$ which is listed in Table 1.   

The H{\sc i} absorption spectrum towards the central component is 
presented in Fig. 2.  H{\sc i} absorption has been detected clearly
towards the inner double of this DDRG. The rms noise in the smoothed spectrum 
is $\sim$0.6 mJy beam$^{-1}$ channel$^{-1}$. The H{\sc i} column density, $N$(H{\sc i}), integrated over the
entire absorption profile using the relation
\begin{equation}
$N$({\rm H{\sc i}})=1.835\times10^{18}\frac{{T}_{s}~\int{\tau(v)dv}}{f_c}~ {\rm cm^{-2}},
\label{eq1}
\end{equation}
where $T_s$, $\tau$ and $f_c$ are the spin temperature, optical depth at a velocity
$v$ and the fraction of background emission covered by the absorber respectively,
is 6.73$\times$10$^{20}$($T_s$/100)(1.0/$f_c$) cm$^{-2}$. The expected value of 
$T_s$ is expected to be significantly greater than 100 K. For example, 
for the warm neutral medium seen in the Galaxy $T_s$ ranges from 5000$-$8000 K 
(Kulkarni \& Heiles 1988). Such high spin temperatures are also expected to arise 
in the proximity of an active nucleus (Bahcall \& Ekers 1969).
The absorption profile consists of a number of components straddling either side
of the optical systemic velocity. The best fit to the present spectrum with 
four Gaussian components is shown in Fig. 2 and the fit parameters
are summarised in Table 2. 
Here v$_{\rm{hel}}$ is the optical heliocentric velocity with an error which is
$\lapp$5 per cent of the full width at half maximum (FWHM). The numbers within 
brackets are the errors on the quoted values. 
More sensitive observations with better spectral resolution
are planned to determine the properties of the individual components more reliably. 

In order to explore any possible relationship between rejuvenation of radio or jet activity
and the occurrence of H{\sc i} we find that most of the DDRGs in the literature have 
weak radio emission in the nuclear region of the host galaxy. 
In the list of DDRGs compiled by Saikia et al. (2006) the two exceptions are 
3C236 and J1247+6723, with the flux density within a few kpc of the nuclear region 
being $\gapp$100 mJy. As noted earlier the DDRG 3C236
shows evidence of star formation and H{\sc i} absorption against a lobe of the inner
radio source (Conway \& Schilizzi 2000; Schilizzi et al. 2001; O'Dea et al. 2001). We have
now reported the detection of H{\sc i} absorption towards the inner double of J1247+6723,
although spectroscopic observations with milliarcsec resolution are required to determine
the association of different absorption features with the different lobes of the mini or inner double.

We have examined the literature for other high-luminosity radio sources with possible evidence 
of rejuvenated nuclear
or radio activity. An interesting case is the radio galaxy 3C293 where the inner double with an 
angular size of $\sim$3.5 arcsec (3 kpc) is misaligned by $\sim$30$^\circ$ from the outer double 
which has an overall
size of $\sim$220 arcsec (192 kpc) and could be classified as a misaligned DDRG. H{\sc i} 
absorption towards the inner double has been observed with  the GMRT and also with higher
resolution against the different components of the inner double using the 
Multi-Element Radio-linked Interferometer Network (MERLIN), Very Large Array and 
Very Long Baseline Interferometric techniques
by Beswick, Pedlar \& Holloway (2002) and Beswick et al. (2004).
Also Morganti et al. (2003) and Emonts et al. (2005) have reported evidence of fast
outflows of neutral hydrogen in 3C293 using the Westerbork Synthesis Radio Telescope.  
Although the fast outflow extending upto $\sim$1000 km s$^{-1}$ is on the blue-shifted
side, absorption components are seen on either side of the optical systemic velocity.
Another interesting source is 3C258 (J1124+1919) where the radio continuum emission is 
dominated by a compact steep spectrum source which was resolved by Sanghera et al. (1995) into a
mini double with no unresolved core component and the lobes separated 
by $\sim$100 milli arcsec (280 pc). 
The source is known to have more extended radio emission (Strom et al. 1990) with a 
largest angular size of
$\sim$60 arcsec (168 kpc). Deep H{\sc i} absorption has been reported recently towards 
the inner double using the Arecibo telescope by Gupta et al. (2006). The spectrum shows
considerable structure with seven components required to fit it satisfactorily. Although
the components appear to be redshifted relative to the systemic velocity a more accurate
measurement of the systemic velocity would be useful. Also while comparing the H{\sc i}
features with the systemic velocity one needs to bear in mind that
the latter could be affected by gas kinematics and one may get different velocity
estimates using different lines (see e.g. Tadhunter et al. 2001; Morganti et al. 2001;
Vermeulen et al. 2006; Gupta \& Saikia 2006a). 

Although the number of galaxies is quite small at present and the statistics need to
be improved, it is interesting to note that
in all four of these galaxies with signs of renewed activity there is occurrence of H{\sc i} 
in absorption. However, there are a few aspects that need to be borne in mind. It is relevant 
to note that there is a high incidence of H{\sc i} absorption in CSS and GPS objects 
(Vermeulen et al. 2003; Pihlstr\"{o}m et al. 2003; Gupta et al. 2006) and
also towards the cores of narrow-line high-luminosity radio galaxies (Morganti et al.
2001; Gupta \& Saikia 2006a). The paucity of detection towards broad-line radio galaxies
(e.g. Morganti et al. 2001; Vermeulen et al. 2003; Gupta et al. 2006) is consistent with
the unification paradigm (e.g. Barthel 1989) with H{\sc i} absorption being largely 
due to disks around the AGN.  The broad-line objects are
expected to be inclined at smaller angles to the line of sight so that the
disk is unlikely to produce significant H{\sc i} absorption. These results
are also consistent with the higher detection rate of
21-cm H{\sc i} absorption towards radio galaxies compared with quasars (Vermeulen et
al. 2003; Gupta et al. 2006). In addition, for a sample of CSS and GPS sources
Gupta \&  Saikia (2006b) find the dependence
of  $N$(H{\sc i}) on the degree of core prominence to be consistent with the H{\sc i}
having a disk-like distribution, with the source sizes being larger than the scale
size of the disk. 

It is interesting to enquire whether in rejuvenated radio galaxies there is likely to
be a higher incidence of H{\sc i} gas in addition to the the circumnuclear disk. At
present although the number of such objects is small the incidence of H{\sc i} absorption 
in rejuvenated powerful radio galaxies appears larger than either CSS or GPS objects. 
For a large sample of 96 radio sources with H{\sc i} information, 
the detection rate for objects $\lapp$15 kpc is only $\sim$35 per cent with this number
increasing to $\sim$45 per cent for the compact GPS objects. For a more reliable 
comparision one requires a larger sample of H{\sc i} observations towards the nuclear 
regions of rejuvenated radio galaxies.  Information on the location of the H{\sc i} gas may be
found from high-resolution observations.
In the case of 3C236, Conway \& Schilizzi (2000) have interpreted the H{\sc i} absorption as
a jet-ISM interaction while in 3C293, Beswick et al. (2002, 2004) find H{\sc i} absorption
towards most components of the inner double. These features do not appear to be due to a
simple disk geometry. While 3C258 and J1247+6723 need to be observed with milliarcsec-scale 
resolution to determine the location of the absorbing clouds, the present data suggest 
that there could be a close relationship between the occurrence of H{\sc i} absorption 
due to gas clouds in the ISM and evidence of renewed activity in powerful radio galaxies.  

\section{Summary}
We have reported the detection of 21-cm absorption towards the inner double of the 
DDRG J1247+6723 using the GMRT. The absorption profile is best fitted by four
components located on either side of the optical systemic velocity, although more sensitive
observations of higher spectral resolution would be useful to determine their 
properties more reliably. 
The total absorbing neutral hydrogen column density of the gas is estimated to be
$N$(H{\sc i})=6.73$\times$10$^{20}$($T_s$/100)($f_c$/1.0)$^{-1}$ cm$^{-2}$. 

Using a small sample of four powerful radio galaxies with evidence of renewed jet activity  we discuss
the possible close relationship between signs of renewed activity and detection of H{\sc i}. 
Although the sample is small the detection rate appears higher than CSS and GPS objects, but
this needs further investigation using a larger sample of rejuvenated radio galaxies with  H{\sc i} 
information. High-resolution spectroscopic observations of two of the objects, 3C236 and 3C293, 
suggest that clouds in the ISM play an important role, while no information is available for
the other two, 3C258 and J1247+6723.  These results are consistent with the possibility of 
a close relationship between the occurrence of H{\sc i} absorption
due to gas clouds in the ISM and evidence of renewed activity in powerful radio galaxies.
Some of these gas clouds may be responsible for `feeding' the 
supermassive black hole with a fresh supply of gas to rejuvenate the nuclear radio
activity. 

\section*{Acknowledgments} 
We thank the reviewer Rob Beswick and the editor for their helpful comments and very prompt 
refereeing of the paper.  We thank our staff for help with the observations.
The Giant Metrewave Radio Telescope is a national facility operated by the National Centre 
for Radio Astrophysics of the Tata Institute of Fundamental Research.
This research has made use of the NASA/IPAC extragalactic database (NED)
which is operated by the Jet Propulsion Laboratory, Caltech, under contract
with the National Aeronautics and Space Administration. We thank numerous contributors
to the GNU/Linux group. 

{}

\end{document}